\documentclass[twocolumn,showpacs,preprintnumbers,amsmath,amssymb]{revtex4}
\usepackage{graphicx}
\usepackage{dcolumn}
\usepackage{bm}
\usepackage{color}

\begin{document}
\preprint{APS/123-QED}

\title{The Smectic $A$ to $C$ Phase Transition in Isotropic
Disordered Environments}
\author{Leiming Chen}
\address{College of Science, The China University of Mining and Technology, Xuzhou Jiangsu, 221116, P. R. China}
\author{John Toner}
\affiliation{Department of Physics and Institute of Theoretical
Science, University of Oregon, Eugene, OR 97403}
\date{\today}
\begin{abstract}
We study theoretically the smectic $A$ to $C$ phase transition in isotropic disordered environments. Surprisingly, we find that, as in the clean smectic $A$ to $C$ phase transition, smectic layer fluctuations do not affect the nature of the transition, in spite of the fact that they are much stronger in the presence of the disorder. As a result, we find that the universality class of the transition is that of  the ``Random field $XY$ model'' ($RFXY$).
\end{abstract} 
\maketitle
The effect of quenched disorder on condensed matter systems
has been  widely studied for many years\cite{Harris,Geoff,Aharony}, both for practical reasons
(since  disorder is always present in real
systems) and fundamental ones. 
Disorder can destroy many types of long ranged order (e.g., ferromagnetic order in systems with quenched random fields \cite{RFferro}), and it can radically change the critical behavior of many phase transitions\cite{Aharony}.

Such effects have been found in, e.g., superconductors\cite{SC}, charge density waves\cite{CDW, CDWRT}, Josephson junction arrays\cite{JJunc}, superfluid helium in aerogel\cite{Helium}, and ferromagnetic superconductors\cite{SCferro}.

Some of the most novel and dramatic effects of quenched disorder are found in liquid crystals confined in random porous media\cite{LC,RT}. These intriguing systems exhibit a variety of exotic ``Bragg Glass" phases. They also undergo unique types of phase transitions\cite{CT}, one of which, the Smectic $A$ to Smectic $C$ (hereafter, $AC$) transition\cite{GP,ACexp}, is the subject of this paper.

In the high temperature phase (the ``$A$" phase),  the nematic
director $\hat{n}$
(which points along the axis of alignment of the constituent long molecules that make up the smectic material), and the normal $\hat{N}$ to the smectic layers , are parallel.  In the low temperature phase (the ``$C$" phase),
$\hat{n}$ and $\hat{N}$ tilt away from each other.

The $AC$ transition in clean systems  was first considered by deGennes\cite{deGennesAC},
who showed that, if
fluctuations of  the smectic layers could be neglected,
the $AC$ transition should be in the universality class of the  ferromagnetic
$XY$ model\cite{Lubensky}.

The effect of layer fluctuations on this result  was considered later by Grinstein and Pelcovits \cite{GP}, who showed that their effect on the $AC$ transition  {\it can}, in fact, be neglected, and that, therefore,
the $AC$ transition in clean systems {\it is} $XY$-like.

Unfortunately, for reasons not yet well understood, the critical region of the $AC$ transition in clean systems appears to be extremely small; as a result, most experimental systems exhibit a transition that is accurately described by mean-field theory\cite{MF}. As a result, no definitive experimental test of the above predictions has yet been made.

Recently the nature of the $AC$ transition has been studied for a liquid crystal confined in
uniaxial\cite{CT} and biaxial\cite{CT2} disordered environments.
In
these systems,
the anisotropy essentially freezes the direction of the molecular axes, and
the $AC$ transition
can be described entirely in terms of the configuration of the smectic layers \cite{CT, CT2}.


In an {\it isotropic } quenched random environment (which can be realized most simply by putting the smectic in aerogel\cite{BirAC}), which we study in this paper,
the problem is in many ways more difficult, since now both fluctuations of the molecular
direction and those of the layers must be addressed. Indeed, it is not even obvious that the two phases between which the transition we wish to study occurs even {\it exist} in $d=3$; the stability of the $A$ phase in the presence of even arbitrarily weak disorder remains an open question
both  theoretically\cite{RT} , and experimentally\cite{LC}. Presumably, similar issues arise with the $C$ phase.

However, if we {\it assume} that both the $A$ and $C$ phases {\it are} stable, then we are able to completely determine the nature of the transition between them.
We find that,  if this stability assumption {\it is} correct, the layer fluctuations do {\it not} affect the universality class of this transition, which proves to be just that of the random field $XY$  model\cite{d->d-2,Dan Fisher RFXY}.

This implies a substantial quantitative change in the universal critical exponents from their values in the clean problem. It is known\cite{d->d-2} that the coefficients in the $\epsilon=6-d$ expansion for the critical exponents of the random field $XY$ model are exactly the same as those for the $\epsilon=4-d$ expansion of the clean (i.e., no random field) problem. However, since $\epsilon=3$ in the physical case $d=3$ for the random field problem, the $\epsilon$-expansion is not quantitatively reliable. It is clear, however, that the exponents will be quite different from those for the clean $XY$ model, as even the first order in  $\epsilon$ terms  change by a factor of 3.

From a quantitative standpoint, the most useful feature of our result is that it connects the exponents of the $AC$ transition in an isotropic disordered environment to those of a random field $XY$ model, as can be experimentally realized in, e.g., anti-ferromagnets with substitutional disorder\cite{birgRF}.

The remainder of this paper is devoted to demonstrating that the $AC$ transition in the presence of isotropic disorder is in the random field $XY$ universality class.




Our starting model  is a  modification of the model for  clean smectics near a Smectic A-Smectic C transition\cite{GP}, the
Hamiltonian $H=H_u+H_c+H_{uc}$ for which consists of three parts:
\begin{eqnarray}
H_u&=&{1\over 2}\int d^dr \left[K(\nabla^2_{\perp}u)^2 + B\left(\partial_z u-{1\over 2}\left|\vec{\nabla}u\right|^2\right)^2\right]
\nonumber,\\
H_c&=&{1\over 2}\int d^dr  \left[K_1\left(\vec{\nabla}\cdot\vec{c}\right)^2+K_2\left(\vec{\nabla}\times\vec{c}\right)^2\right.\nonumber\\
   &&\left.+K_3\left(\partial\vec{c}\over\partial z\right)^2+D c^2+2vc^4\right],\nonumber\\
H_{uc}&=&{1\over 2}\int d^dr  \left[g_1c^2\left(\partial_z u-{1\over 2}|\vec{\nabla}u|^2\right) +g_2(\nabla_{\perp}^2u)\times\right.\nonumber\\
&&\left.(\vec{\nabla}\cdot\vec{c})
+g_3\left(\partial\vec{c}\over\partial z\right)\cdot\left(\partial_z\vec{\nabla}_{\perp}u\right)+D(\vec{c}\cdot\vec{\nabla}_{\perp}u)^2\right]\nonumber,
\end{eqnarray}
where we have defined the direction parallel to the averaged layer normal in the $A$ phase as the $\hat{z}$-axis, and
the plane perpendicular to $\hat{z}$ as $\perp$. Here $\vec{c}$, which is roughly the projection of $\hat{n}$ onto the layers, is the order parameter for the transition.
More precisely, it has only two non-zero components (i.e., $ c_z(\vec{r})=0$), given by
\begin{eqnarray}
 c^{\perp}_i(\vec{r})&=&\left[\delta_{ij}-N_i(\vec{r}) N_j(\vec{r})\right]n_j(\vec{r}), \ \ i\in\perp,
\end{eqnarray}
where $\hat{N}(\vec{r})$ denotes the unit vector along the layer normal, given by $\hat{N}={\hat{z}-\vec{\nabla}u\over \mid\hat{z}-\vec{\nabla}u\mid}$.
Note that all terms in the Hamiltonian are rotation invariant. This is because the environment is isotropic and rotating
the smectic does not cost energy. The pieces $H_u$ and $H_c$ are, respectively, just the elastic energies for smectic layer fluctuations and  molecular reorientations,  while $H_{uc}$
 couples $u$ and $\vec{c}$.

The fourth term in $H_c$ and the last term in $H_{uc}$ , which come from a term  $D(T)\mid\hat{N}-\hat{n}\mid^2$, induce the $AC$ transition via a sign change in the temperature $T$-dependent coefficient $D(T)$. For  $T>T_{AC}$, $D>0$, and  the free energy is minimized at $\vec{c}=\vec{0}$, so the system is in the $A$ phase. For $T<T_{AC}$, $D<0$, and  the free energy is minimized at $\vec{c}\ne\vec{0}$, so the system is in the $C$ phase.



Now we include disorder.
The  aerogel exerts a variety of random forces on the  molecular axes and the  smectic layers\cite{RT,CT}; the most important of
them
are\cite{RT, CT} the  ``random tilt fields", which tend to align the local molecules  and the local normals with the random aerogel strands. The contribution of these random effects to the free energy can be written as\cite{RT,CT}
\begin{eqnarray}
 \int d^dr \left[\vec{h}(\vec{r})\cdot\vec{\nabla}_{\perp}u+\vec{h}^c(\vec{r})\cdot\vec{c}\right],\label{disorder}
\end{eqnarray}
where the quenched random fields $\vec{h}(\vec{r})$ and  $\vec{h}^c(\vec{r})$ are taken to have  Gaussian distributions of  zero mean, with anisotropic short-ranged correlations:
\begin{eqnarray}
 \overline{h_i(\vec{r})h_j(\vec{r}\,')}=\Delta\delta^{\perp}_{ij}\delta^d(\vec{r}-\vec{r}\,'), \\ \overline{h^c_i(\vec{r})h^c_j(\vec{r}\,')}=\Delta_c\delta^{\perp}_{ij}\delta^d(\vec{r}-\vec{r}\,'),\\
\overline{h^c_i(\vec{r})h_j(\vec{r}\,')}=\Delta'\delta^{\perp}_{ij}\delta^d(\vec{r}-\vec{r}\,').
\end{eqnarray}
The first term in equation (\ref{disorder}) has been treated  in the earlier work\cite{RT} on the smectic $A$ phase in isotropic disordered
environments, where it leads to strong power-law anomalous\cite{RT}. The second term is just the random
field disorder present in the $RFXY$ model\cite{RFferro, Dan Fisher RFXY}.

To cope with the quenched disorder we employ the replica trick \cite{Geoff}. We assume that the free energy of the system for a specific
realization of the disorder is the same as that  averaged over many realizations. To calculate the averaged
free energy $\overline{F}=\overline{\ln{Z}}$, where $Z$ is the partition function, we use
the mathematical identity $\ln{Z}=\lim_{n\to 0}{Z^n-1\over n}$. When calculating $\overline{Z^n}$, we
can first compute the average over the random fields $\vec{h}(\vec{r})$,
whose statistics have been given earlier. Implementing this procedure gives a replicated Hamiltonian $H^r=H^r_u+H^r_c+H^r_{uc}$
with the effect of the random fields transformed into couplings between $n$ replicated fields, with the limit $n\rightarrow 0$ corresponding to the original quenched disorder problem: \begin{eqnarray}
H_u^r&=&{1\over 2}\int d^dr \sum_{\alpha=1}^{n} \left[B\left(\partial_z u_{\alpha}-{1\over 2}\left|\vec{\nabla}u_{\alpha}\right|^2\right)^2\right.\nonumber\\
&&\left.+K(\nabla^2_{\perp}u_{\alpha})^2\right]-{\Delta\over 2k_BT}\int d^dr \sum_{\alpha,\beta=1}^{n} \vec{\nabla}_{\perp}u_{\alpha}\cdot\vec{\nabla}_{\perp}u_{\beta},
\nonumber\\
\label{Hur}\\
H_c^r&=&{1\over 2}\int d^dr  \sum_{\alpha=1}^n\left[K_1\left(\vec{\nabla}\cdot\vec{c}_{\alpha}\right)^2+K_2\left(\vec{\nabla}_\perp\times\vec{c}_{\alpha}\right)^2\right.\nonumber\\
&&\left.+K_3\left(\partial\vec{c}_{\alpha}\over\partial z\right)^2+D c_{\alpha}^2+2vc_{\alpha}^4\right]\nonumber\\
&&-{\Delta_c\over 2k_BT}\int d^dr \sum_{\alpha,\beta=1}^{n} \vec{c}_{\alpha}\cdot\vec{c}_{\beta},\label{RFXY}\\
H_{uc}^r&=&{1\over 2}\int d^dr \left[\sum_{\alpha=1 }^n \left(g_1c_{\alpha}^2\left(\partial_z u_{\alpha}-{1\over 2}|\vec{\nabla}u_{\alpha}|^2\right)\right.\right.\nonumber\\
&&\left. \left.+g_2(\nabla_{\perp}^2u_{\alpha})(\vec{\nabla}\cdot\vec{c}_{\alpha})+g_3\left(\partial\vec{c}_\alpha\over\partial z\right)\cdot\left(\partial_z\vec{\nabla}_{\perp}u_\alpha\right)\right.\right.\nonumber\\
&&\left.\left.+D(\vec{c}_\alpha\cdot\vec{\nabla}_{\perp}u_{\alpha})^2\right)
\right].
\label{Hucr}
\end{eqnarray}

If we set $u_{\alpha}=0$, the entire Hamiltonian reduces to Eq. (\ref{RFXY}),
which reduces to the $RFXY$ model if $K_1=K_2=K_3$. An RG
analysis shows that departures from this  ``one constant approximation" (i.e., $K_{1,2,3}=K$)
are irrelevant\cite{foot3}; hence, in the absence of the $u$ field,
the transition is in the $RFXY$ universality class.

The piece $H_u^r$ Eq. (\ref{Hur}) of $H$ which involves $u$ alone is precisely the model for
smectics $A$ in isotropic aerogel studied in reference\cite{RT}. From the analysis of that reference, we
know that the critical dimension of Eq. (\ref{Hur}), below which the anharmonic terms in Eq.
(\ref{Hur}) become important, is $5$. On the other hand,  the critical dimension of $H_c^r$ Eq.
(\ref{RFXY}) is well known\cite{RFferro, Dan Fisher RFXY, d->d-2} to be $6$. Because of this discrepancy
between the two critical dimensions,  a standard $\epsilon$-expansion study of the entire model Eqs.
(\ref{Hur}-\ref{Hucr})" is impossible.
Our solution to this quandry is to integrate out {\it only} the $u_\alpha$ fields perturbatively in a momentum shell RG approach, which is controlled in an $\epsilon=5-d$-expansion, to obtain an effective model that only involves $\vec{c}_{\alpha}$.
While unorthodox, this approach is very much in the spirit of more conventional RG's: we are performing a partial trace over some degrees of freedom to obtain a more tractable Hamiltonian in terms of the degrees of freedom remaining after the trace.

The momentum shell RG procedure consists of tracing over the short wavelength Fourier modes
of $u_{\alpha}(\vec{r})$ followed by a rescaling of the length.
We initially restrict wavevectors to lie in a bounded Brillouin zone which we take to be the infinite cylinder $0<\mid\vec{q}_{\perp}\mid<\Lambda$,
$-\infty<q_z<\infty$, where $\Lambda\sim 1/a$ is an
ultra-violet cutoff, and $a$ is the smectic layer spacing. The displacement field $u_{\alpha}(\vec{r})$
is separated into high and low wave vector parts
$u_{\alpha}(\vec{r})=u_{\alpha}^<(\vec{r})+u_{\alpha}^>(\vec{r})$,
where $u_{\alpha}^>(\vec{r})$ has support in the thin wave vector shell $\Lambda
e^{-d\ell}<\mid\vec{q}_{\perp}\mid<\Lambda$, $-\infty<q_z<\infty$. Here $d\ell$ is a differential parameter $d\ell\ll1$. We first integrate out $u^>_\alpha(\vec{r})$. This integration is done perturbatively in
 the anharmonic terms in
$H$ Eqns. (\ref{Hur})-(\ref{Hucr}).  After this
perturbative step, we anisotropically rescale lengths,
with $\vec{r}_{\perp}=\vec{r}^{~\prime}_{\perp}e^{\ell}$, $r_z=r^\prime_z e^{\omega\ell}$, so as to restore the UV cutoff back to $\Lambda$. This is then followed
by rescaling the long wave length part of the field with $u_{\alpha}^<(\vec{r})=u'_{\alpha}
(\vec{r'})e^{\chi \ell}$. The underlying rotational invariance insures that the graphical
corrections preserve the rotationally invariant operator
$\partial_z u_{\alpha}-{1\over 2}\left(\vec{\nabla}u_{\alpha}\right)^2$ renormalizing it as a whole. It is therefore
convenient  to choose the dimensional
rescaling that also preserves this operator, which is $ \chi=2-\omega$.

After this procedure we obtain the following RG flow equations to
one-loop order, ignoring the term $D \left(\vec{c}_{\alpha}\cdot\vec{\nabla}_{\perp}u_{\alpha}\right)^2$, since we are interested in the critical point where $D$ vanishes:
\begin{eqnarray}
 {dB\over d\ell}&=&\left(d+3-3\omega-{3\over 16}g\right)B,\label{Bflow}\\
 {dK\over d\ell}&=&\left(d-1-\omega+{1\over 32}g\right)K,\label{Kflow}\\
 {d\Delta\over d\ell}&=&\left(d+1-\omega+{1\over 64}g\right)\Delta,\label{Deltaflow}\\
 {dg_1\over d\ell}&=&\left(d+1-\omega-{3\over 16}g\right)g_1,\label{g1flow}\\
 {dv\over d\ell}&=&\left(d-1+\omega-{3g\over 128}{g_1^2\over Bv}\right)v.\label{vflow}
\end{eqnarray}
where $g$ is a dimensionless coupling:
\begin{eqnarray}
g\equiv \Delta \left(B\over K^5\right)^{1\over 2}C_{d-1}\Lambda^{d-5},
\end{eqnarray}
where $C_d$ is the surface area of a $d$-dimensional sphere with radius one divided by $(2\pi)^d$.

\begin{figure}
 \includegraphics[width=0.4\textwidth]{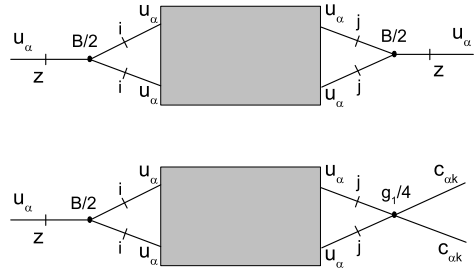}
 \caption{\label{fig: FGraph}Schematic representation of all Feynman graphs that renormalize $B$  (top diagram) and $g_1$ (bottom diagram). In each case the gray box  represents all possible ways of connecting the portions of the Feynman graphs shown. These are identical for both graphs; as a result, the graphical corrections to $B$ and $g_1$ obey $(dB/dl)_{graph}=(dg_1/dl)_{graph}  (g_1/B)$. This in turn implies that the anomalous elasticity for $g_1$ (see text) is identical, up to a multiplicative constant, to that for $B$.
}
\end{figure}

Note that the graphical corrections inside the parenthesis in Eqs. (\ref{Bflow}) and (\ref{g1flow}) are the same. This is {\it not} just an
an approximation to one-loop order, but {\it exact} to arbitrary loop order. This can be easily understood by analyzing the structures of
the Feynman graphs. In Fig. \ref{fig: FGraph} the upper graph summaries all the possible graphical corrections to $(\partial_zu_{\alpha})\mid\vec{\nabla}_{\perp}u_{\alpha}\mid^2$; the lower one does for $(\partial_zu_{\alpha})c_{\alpha}^2$. The parts inside the two square boxes are the same no matter how complicated they are and how many loops they have.

There are no graphical corrections to $(\nabla_{\perp}^2u_{\alpha})(\vec{\nabla}\cdot\vec{c}_{\alpha})$, which is also {\it exact} to arbitrary-loop order.
This is because both terms have one power of $c_{\alpha}$ while all anharmonic terms have even powers of $c_{\alpha}$. Therefore,
under renormalization both $g_{2,3}$ flow only as a result of length and field rescaling.

The recursion relations for $B$, $K$, and $\Delta$ are {\it identical} with those found for a smectic $A$ in an isotropic disordered medium in reference\cite{RT}. This is also exact to all orders, since we have not, in our unusual approach, integrated out the $\vec{c}$ fluctuations.
This means that all of the results obtained in \cite{RT} for the long-wavelength behavior of these quantities also hold here. We will also make use of {\it this} fact later.

\begin{figure}
 \includegraphics[width=0.4\textwidth]{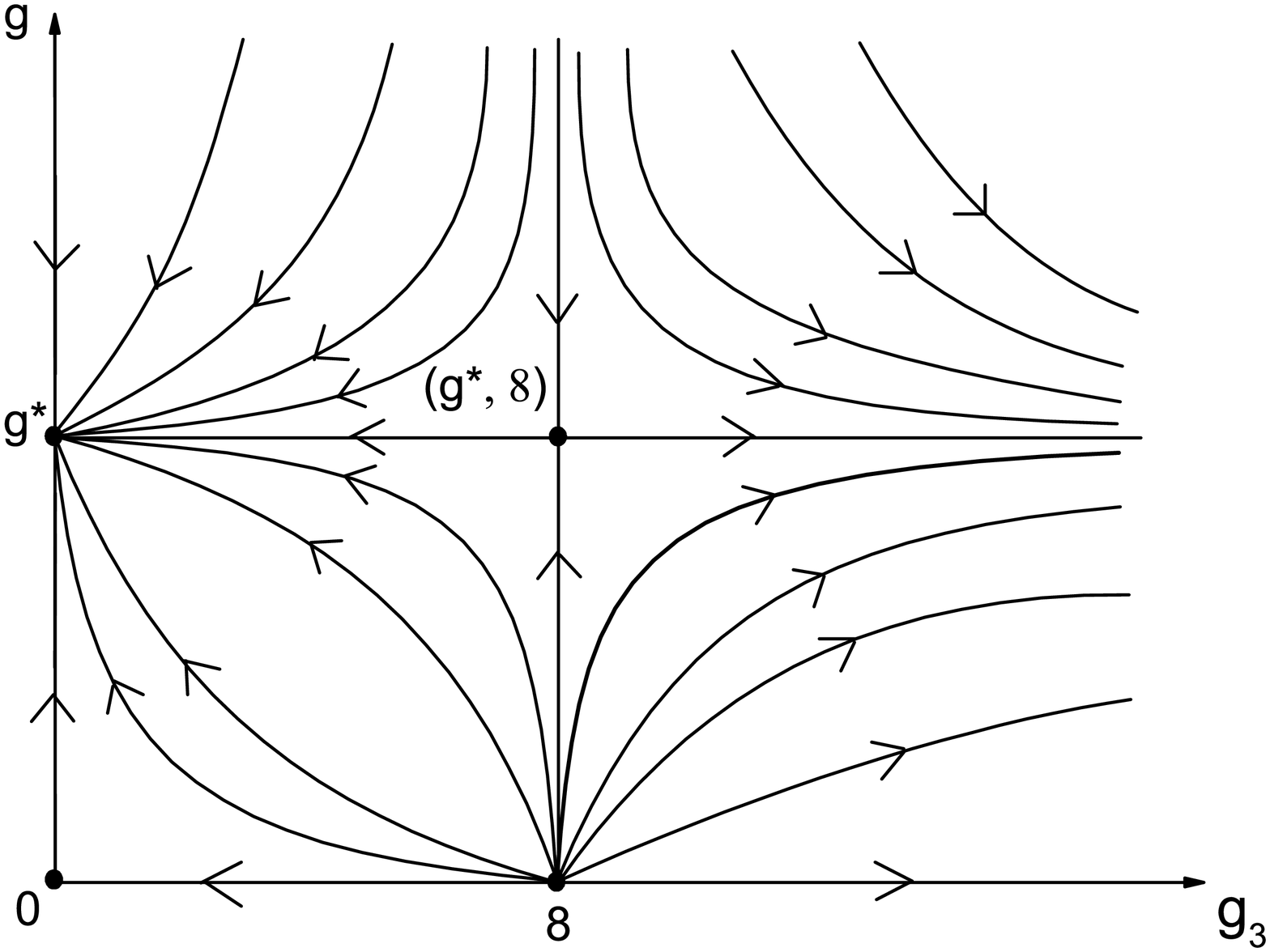}
 \caption{\label{fig: RGFlow}RG flows of the dimensionless couplings $g$ and $g_3$ from equations (\ref{gflow}) and (\ref{g4flow}). All initial models starting to the left of the stability limit $g_3=8$ flow into the $g=g^*$, $g_3=0$ fixed point, which therefore controls the $AC$ transition. All models starting to the right of the stability limit are unstable.}
\end{figure}

To analyze these flow equations we introduce an additional dimensionless coupling: $
 g_3 \equiv {g_1^2\over Bv}$.
Combining flow Eqs. (\ref{Bflow}-\ref{vflow}) with the definitions of $g$ and $g_3$ we find
\begin{eqnarray}
{dg\over d\ell}&=&\epsilon g-{5\over 32}g^2,\label{gflow}\\
{dg_3\over d\ell}&=&{3g\over 128}\left(-8+g_3\right)g_3\label{g4flow},
\end{eqnarray}
where $\epsilon=5-d$.
These flow equations have four fixed points: $
g^*=0\ \ \mbox{or}\ \ {32\over 5}\epsilon,
g_3^*=0\ \ \mbox{or}\ \ 8$.
The RG flows of $g$ and $g_3$ around these fixed points are illustrated in Fig. \ref{fig: RGFlow}.
Note that $g_3^*=8$ corresponds to the stability limit of the system.
Linearizing Eqs. (\ref{gflow}, \ref{g4flow}) around the only stable fixed point $g^*={32\over 5}\epsilon$, $g_3^*=0$, we find the graphical corrections to $v$ vanish exponentially as  $\ell\to\infty$. This implies that integrating out $u_{\alpha}$ only gives a finite correction to $v$, even at arbitrarily long wavelengths. Hence, these corrections to $v$ coming from the $u_\alpha$ fields do not affect the nature of the $AC$ transition.

During each RG cycle the integration over $u_{\alpha}^>$ also generates terms which do not exist in $H_c^r$. The most relevant ones
are produced in the second cumulant by $(\partial_z u_{\alpha})c_{\alpha}^2$ and $(\nabla_{\perp}^2 u_{\alpha})(\vec{\nabla}\cdot\vec{c}_{\alpha})$.
Elementary power counting shows that the terms generated by $\left(\partial\vec{c}_{\alpha}\over\partial z\right)\cdot\left(\partial_z\vec{\nabla}_{\perp}u_{\alpha}\right)$ are less relevant.

We'll now show that these terms also do not affect the nature of the $AC$ transition.
We start with the terms generated by $(\partial_z u_{\alpha})c_{\alpha}^2$:
\begin{eqnarray}
 &&\sum_{\alpha,\beta}^n \sum_{\vec{q}_{1,2}\vec{k}}g_1^2(\vec{k})\left[{k_BTk_z^2 G(\vec{k})}\delta_{\alpha\beta}+{\Delta(\vec{k})k_z^2 k_{\perp}^2G^2(\vec{k})}\right]\nonumber\\
 &&\times c_{\alpha,i}(\vec{q}_1)c_{\alpha,i}(-\vec{q}_1+\vec{k})
 c_{\beta,i}(\vec{q}_2)c_{\beta,i}(-\vec{q}_2-\vec{k})
\label{Newterm1}
\end{eqnarray}
where $G(\vec{k})\equiv 1/[B(\vec{k})k_z^2+K(\vec{k})k_{\perp}^4]$. The $\vec{k}$-dependences of $B$, $K$, $\Delta$, and $g_1$ arise due to the  the nonzero graphical corrections in the recursion relations Eqs. (\ref{Bflow}-\ref{g1flow}).
Because, as mentioned earlier, Eqs. (\ref{Bflow}-\ref{Deltaflow}) are {\it identical}, to all orders, with those for a smectic $A$ in an isotropic disordered environment, we can simply use the results of \cite{RT} for the wavevector dependences of these quantities.
Furthermore, since, as noted earlier, there is an exact relation between the
renormalization of $g_1$ and that of $B$, the wavevector dependence of $g_1$ is identical to that
of $B$, up to an overall multiplicative constant.

Using the just noted connections to the work of \cite{RT}, we can simply quote  $\vec{k}$-dependences of $B$, $K$, $\Delta$, and $g_1$:
\begin{eqnarray}
 B(\vec{k}), g_1(\vec{k})\propto\left\{
 \begin{array}{ll}
  k_{\perp}^{\eta_B}, &k_z\ll k_{\perp}^{\zeta},\\
  k_z^{\eta_B/\zeta}, &k_z\gg k_{\perp}^{\zeta},
 \end{array}
 \right.
 \label{B}
\end{eqnarray}
\begin{eqnarray}
 K(\vec{k}), \Delta(\vec{k})\propto\left\{
 \begin{array}{ll}
  k_{\perp}^{-\eta_{K, \Delta}}, &k_z\ll k_{\perp}^{\zeta},\\
  k_z^{-\eta_{K,\Delta}/\zeta}, &k_z\gg k_{\perp}^{\zeta},
 \end{array}
 \right.
 \label{K}
\end{eqnarray}
where the anisotropy scaling exponent $
 \zeta=2-{{\eta_B+\eta_K}\over 2}$, and $\eta_{B,K,\Delta}>0$.
Another result of \cite{RT} is that the exponents $\eta_{B,K,\Delta}$ are not fully independent, but connected by the {\it exact} scaling relation:
\begin{eqnarray}
 5-d+\eta_{\Delta}={\eta_B\over 2}+{5\over 2}\eta_K,\label{Scaling1}
\end{eqnarray}
which is implied by the fact that $g$ flows to a nonzero stable fixed point \cite{RT}. Furthermore,
there are certain bonds on the values of $\eta_{B,K}$ that must be satisfied in order for the smectic $A$ phase in an isotropic random environment  to be stable, which is a prerequisite condition for the existence of a sharp smectic $A$-$C$ transition \cite{RT} in such environments. It is only meaningful within these bounds to discuss the relevance of the terms in formula (\ref{Newterm1}). These bounds are
\begin{eqnarray}
 \eta_K+\eta_B<2,~~\eta_K<1,~~ \eta_B+5\eta_K>4.\label{Bound1}
\end{eqnarray}
The first two bounds come from the requirement
of long-ranged orientational order and the condition
for dislocations to remain confined, respectively. The third bound is obtained by combining $\eta_{\Delta}>0$ with the exact scaling relation
(\ref{Scaling1}) in $d=3$.

Using expressions (\ref{B}, \ref{K}) we can write equation (\ref{Newterm1}) in a scaling form:
\begin{eqnarray}
 &&\sum_{\alpha,\beta}^n \sum_{\vec{q}_{1,2}\vec{k}}\left[k_{\perp}^{\eta_B}f_1\left(k_z\over k_{\perp}^{\zeta}\right)\delta_{\alpha\beta}+k_{\perp}^{(\eta_B-3\eta_K)/2}f_2\left(k_z\over k_{\perp}^{\zeta}\right)\right]\nonumber\\
 &&\times c_{\alpha,i}(\vec{q}_1)c_{\alpha,i}(-\vec{q}_1+\vec{k})
 c_{\beta,i}(\vec{q}_2)c_{\beta,i}(-\vec{q}_2-\vec{k}),~~~\label{Scalingform}
\end{eqnarray}
where $f_{1,2}(x)$ are scaling functions.
Clearly, as $\vec{k}\to\vec{0}$ the replica-diagonal term (i.e., the one which contains $\delta_{\alpha\beta}$) in (\ref{Scalingform}) is irrelevant compared to the quartic ($v$) term in $H_c^r$, since its coefficient vanishes like $k_{\perp}^{\eta_B}$.

 To decide whether the off-diagonal piece is relevant, we treat it as a perturbation and calculate its contributions to $D$:
\begin{eqnarray}
 \delta D&=&\int d^dk \ \ k_{\perp}^{(\eta_B-3\eta_K)/2}f_2\left(k_z\over k_{\perp}^{\zeta}\right){1\over ck^2+D}\nonumber\\
         &=&\int d^dk \ \ k_{\perp}^{(\eta_B-3\eta_K)/2}f_2\left(k_z\over k_{\perp}^{\zeta}\right){1\over ck^2}\left(1-{D\over ck^2}\right).\nonumber
\end{eqnarray}
It is readily  shown that this integral converges for  $d$ near $6$ if the exponents $\eta_{B, K}$ satisfy  the bounds (\ref{Bound1}). Therefore, this off-diagonal piece is also irrelevant.

Now we discuss the terms generated by $(\nabla_{\perp}^2 u_{\alpha})(\vec{\nabla}\cdot\vec{c}_{\alpha})$, which also have a diagonal and an off-diagonal part:
\begin{eqnarray}
\sum_{\alpha,\beta}^n \sum_{\vec{q}}&&g_2^2\left[k_BT{q_{\perp}^4 G(\vec{q})}\delta_{\alpha\beta}+{\Delta(\vec{q}) q_{\perp}^6G^2(\vec{q})}\right]\times\nonumber\\
 &&q_iq_j c_{\alpha,i}(\vec{q})c_{\beta,j}(-\vec{q}).\label{last}
\end{eqnarray}
Here, unlike $g_1$, $g_2$ has {\it no} dependence on $\vec{q}$ since there are no graphical corrections to $(\nabla_{\perp}^2u_{\alpha})(\vec{\nabla}\cdot\vec{c}_{\alpha})$. Again we can rewrite Eq. (\ref{last}) in a scaling form:
\begin{eqnarray}
\sum_{\alpha,\beta}^n \sum_{\vec{q}}&&g_2^2\left[q_{\perp}^{\eta_K}f_3\left(q_z\over q_{\perp}^{\zeta}\right)\delta_{\alpha\beta}+q_{\perp}^{-\left(\eta_B+3\eta_K\right)/2}f_4\left(q_z\over q_{\perp}^{\zeta}\right)\right]\nonumber\\
 &&\times q_iq_j c_{\alpha,i}(\vec{q})c_{\beta,j}(-\vec{q}),\label{}
\end{eqnarray}
where $f_{3,4}(x)$ are scaling functions similar to $f_{1,2}(x)$. Clearly, both terms are subdominant to the
quadratic terms in $H_c^r$ as $q\to 0$ provided that $\eta_{B,K}$ are within the stability bounds.

Therefore, we conclude that integrating out $u_{\alpha}$ only gives minor corrections to $H_c^r$, which do not affect the nature of the transition. Therefore,  the universality class of the transition is just that of the random field $XY$ model,
as it would be were the full Hamiltonian just $H_c^r$.

In summary, in this paper we've  studied the smectic $A$ to $C$ phase transition in isotropic disordered environment.
Our analysis shows that if the smectic phases are stable against fluctuations and unbinding of dislocations,
the universality class of the transition is that of the ``Random Field $XY$ Model''. Surprisingly, in spite of the fact
that the smectic layer fluctuations are large due to the disorder, they have no effect on the nature of the transition;
that is, if the layers can be frozen by some experimental means the universality class of the transition still
remains the same. During this study we developed a``partial renormalization group" strategy which proves to be very successful.
We expect this strategy to be useful in dealing with many problems with anharmonic Hamiltonians which involve multiple fields
with different critical dimensions.

We thank G. Grinstein for valuable discussions, and for leading JT to the ice cream place under the Brooklyn Bridge. We're both  also grateful to the MPIPKS, Dresden, where a portion of this work was done, for their support (financial and otherwise) and hospitality. JT thanks the U.S.  National Science Foundation for their financial support
through awards \# EF-1137815 and 1006171; LC acknowledges support by the National Science Foundation of China (under
Grant No. 11004241) and the Fundamental Research
Funds for the Central Universities (under Grant No.
2010LKWL09). LC also thanks the China Scholarship Fund for supporting his visit to the University of Oregon, where a portion of this work was done. He is also grateful to the University of Oregon's Physics Department and Institute for Theoretical Science for their hospitality.


\begin{thebibliography}{99}
\bibitem{foot3}
The RG analysis is similar to the pure case \cite{GP}.
We find that,  in an $\epsilon=6-d$ expansion,  the eigenvalue $\lambda_{\delta K}$ of this perturbation is $-n\epsilon^2\over {3(n+8)^2}$, where $n$ denotes the number of components of $\vec{c}$.
\end{thebibliography}
\end{document}